\documentclass[11pt]{article}
\usepackage{moriond,epsfig}

\def\Journal#1#2#3#4{{#1} {\bf #2}, #3 (#4)}

\def\NPB{{\em Nucl. Phys.} B}

\def\PRL{\em Phys. Rev. Lett.}
\def\PRD{{\em Phys. Rev.} D}

\def\sst{\scriptscriptstyle}

\def\ra{\rightarrow}

\def\be{\begin{equation}}
\def\ee{\end{equation}}
\def\bea{\begin{eqnarray}}
\def\eea{\end{eqnarray}}

\begin{document}
\hfill{SISSA Ref. 42/2003/EP}
\vspace*{4cm}
\title{LEPTON FLAVOR VIOLATING DECAYS AND QUASI-DEGENERATE NEUTRINOS}

\author{ CARLOS E. YAGUNA }

\address{SISSA-ISAS, Via Beirut 2-4,\\
Trieste  34014, Italy}

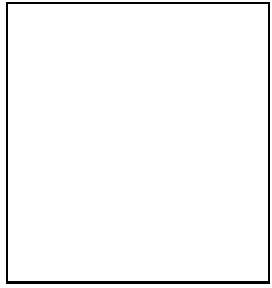
\maketitle\abstracts{We show that the predictions for lepton flavor violating decays, in the case of quasi-degenerate neutrinos, change drastically when we include in the computation the high energy phases present in the matrix of neutrino Yukawa couplings.}

In supersymmetric models with right-handed (RH) neutrinos, lepton flavor violating effects are generated by renormalization group evolution between the unification scale ($M_X$) and the RH neutrino scale ($M_R$) \cite{masiero}. At $M_{X}$ the slepton mass matrix is usually assumed to be diagonal (e.g. in m-SUGRA models), $(m_{\sst L}^2)_{ij}=m_0^2 \delta_{ij}$; between $M_X$ and $M_R$ it acquires non-zero  off-diagonal elements that are given, in the logarithmic approximation, by
\be
(m_{\sst L}^2)_{ij}=-\frac{1}{16 \pi^2}(6+2 a_0^2) m_0^2 (\mathbf{Y_\nu^\dagger Y_\nu})_{ij} \log\frac{M_{X}}{M_R}\,,
\label{log}\ee  
where $\mathbf{Y_\nu}$ is the matrix of neutrino Yukawa couplings.
These off-diagonal elements determine the rate of lepton flavor violating processes. In fact, the branching ratio for the processes $\ell_i\rightarrow \ell_j+\gamma$ is approximately given by  \cite{hisano} 
\be
Br(\ell_i\rightarrow \ell_j+\gamma)\approx \frac{\alpha^3}{G_F^2}\frac{\left|\left(m_{\sst L}^2\right)_{ij}\right|^2}{m_{\sst S}^8}\tan^2\beta\,,
\label{br}\ee
where $m_{\sst S}$ is a typical supersymmetric mass. In contrast with the predictions of the Standard Model, such branchings are generically close to the experimental bounds.

We see from Eqs.(\ref{log}) and (\ref{br}) that, for a given set of soft-breaking parameters, the predictions for LFV decays depend on the quantities $(\mathbf{Y_\nu^\dagger Y_\nu})_{ij}$. If neutrino masses are generated from the see-saw mechanism, the most general neutrino Yukawa coupling can be written as \cite{casas}
\be
\mathbf{Y_\nu}=\frac{1}{v_u}D_M^{\sst 1/2} \mathbf{R} D_m^{\sst 1/2}U^\dagger
\ee
where $v_u$ is the VEV of $H_u$, $D_M$ is the diagonal mass matrix of RH neutrinos, $U$ is the mixing matrix of light neutrinos, $D_m$ is the diagonal matrix of light neutrino mass eigenvalues, and $\mathbf{R}$ is an arbitrary complex orthogonal matrix. 

It is usually assumed that $\mathbf{R}$ is real and $D_M=M_R \mathbf{1}$. In that case one gets
\be
(\mathbf{Y_\nu^\dagger Y_\nu})_{21}=\frac{M_R}{v_u^2}\left[U_{22}U^*_{12} (m_2-m_1)+U_{23}U^*_{13} (m_3-m_1)\right]\,,
\label{real}\ee
which is independent of $\mathbf{R}$. Notice that this quantity depends on the masses  only through mass differences. Since for quasi-degenerate neutrinos, $m_i-m_j\ll m_{i,j}$, this means that very strong cancellations are taking place. Besides,  the largest mass difference, $m_3-m_1$, is partially killed by $U_{13}$ --the smallest element in $U$. If, instead,  $\mathbf{R}$ is complex then such cancellations in $(\mathbf{Y_\nu^\dagger Y_\nu})_{21}$ no longer occur and, therefore, we expect a branching that is much larger than the one obtained from (\ref{real}). The dependece of $\mathbf{Y_\nu^\dagger Y_\nu}$ on $\mathbf{R}$, however, does not cancel out and we have to deal with 6 additional unknown parameters.

When the spectrum is degenerate, three of the six parameters in $\mathbf{R}$ can be rotated away. The remaining three enter also into the expression for the CP asymmetry in RH neutrino decays which, in leptogenesis, is responsible for the baryon asymmetry of the Universe.  We use a specific model of leptogenesis to put a constraint on their product and then calculate  the rate for the processes $\ell_i\ra \ell_j+\gamma$ for $\mathbf{R}$ real and complex. We found that the branchings for $\mathbf{R}$ complex are always enhanced by at least two orders of magnitude with respect to the ones  calculated for $\mathbf{R}$ real \cite{yaguna}. Our results are summarized in Table 1.
\vspace{0.4cm}
\begin{table}[h]\begin{center}
\caption{Enhancenment in Br($\ell_i\ra \ell_j+\gamma$). }
\begin{tabular}{|c|c|c|}
\hline
 & \multicolumn{2}{c|}{ \rule[-3mm]{0mm}{8mm}\textbf{Enhancement}}\\ \cline{2-3}
 \raisebox{2.0ex}[0pt]{\textbf{ Process}}& \rule[-2mm]{0mm}{6mm}\hspace{3mm}$U_{13}=0.0$\hspace{3mm} & $U_{13}=0.2$\\
\hline\hline
\rule[-2mm]{0mm}{6mm}$\mu\ra e+\gamma$ & $10^5$ & $10^3$\\
\rule[-2mm]{0mm}{6mm}$\tau\ra \mu+\gamma$ & $10^2$ & $10^2$\\
\rule[-2mm]{0mm}{6mm}$\tau\ra e+\gamma$ & $10^5$ & $10^3$\\
\hline
\end{tabular}  \end{center}
\end{table}

\section*{Acknowledgments}
This work was done in collaboration with S. Pascoli and S. Petcov. I would like to thank the organizers of \emph{Rencontres de Moriond 2003} for the stimulating environment, and SISSA for financial support.

\end{document}